\documentclass[]{IEEEtran}
\usepackage{cite}
\usepackage{amsmath,amssymb,amsfonts}
\usepackage{algorithmic}
\usepackage{graphicx}
\usepackage{textcomp}
\usepackage{xcolor}
\usepackage{multirow}
\usepackage{caption} 
\usepackage{float}
\usepackage{soul}
\usepackage{xcolor}
\usepackage{amsmath}
\usepackage{hyperref}

\usepackage{array}
\newcommand{\PreserveBackslash}[1]{\let\temp=\\#1\let\\=\temp}
\newcolumntype{C}[1]{>{\PreserveBackslash\centering}p{#1}}
\newcolumntype{R}[1]{>{\PreserveBackslash\raggedleft}p{#1}}
\newcolumntype{L}[1]{>{\PreserveBackslash\raggedright}p{#1}}
\usepackage{booktabs}

\usepackage{subcaption}
\hypersetup{
    colorlinks=true,
    allcolors=blue
}

\def\BibTeX{{\rm B\kern-.05em{\sc i\kern-.025em b}\kern-.08em
    T\kern-.1667em\lower.7ex\hbox{E}\kern-.125emX}}
\graphicspath{ {images/} }

\begin{document}
\pagestyle{empty}
\title{AttResDU-Net: Medical Image Segmentation Using Attention-based Residual
Double U-Net} 

 \author{ 
     \IEEEauthorblockN{Akib Mohammed Khan\textsuperscript{*}\thanks{* Contributed Equally}, Alif Ashrafee\textsuperscript{*}, Fahim Shahriar Khan\textsuperscript{*}, Md. Bakhtiar Hasan, and Md. Hasanul Kabir\\}

     \IEEEauthorblockA{\textit{Department of Computer Science and Engineering,\\Islamic University of Technology, Gazipur, Bangladesh\\}}
    
     \IEEEauthorblockA{\{akibmohammed, alifashrafee, fahimshahriar52, bakhtiarhasan, hasanul\}@iut-dhaka.edu}
 }



\maketitle
\thispagestyle{empty}

\begin{abstract}
Manually inspecting polyps from a colonoscopy for colorectal cancer or performing a biopsy on skin lesions for skin cancer are time-consuming, laborious, and complex procedures. Automatic medical image segmentation aims to expedite this diagnosis process. However, numerous challenges exist due to significant variations in the appearance and sizes of objects with no distinct boundaries. This paper proposes an attention-based residual Double U-Net architecture (AttResDU-Net) that improves on the existing medical image segmentation networks. Inspired by the Double U-Net, this architecture incorporates attention gates on the skip connections and residual connections in the convolutional blocks. The attention gates allow the model to retain more relevant spatial information by suppressing irrelevant feature representation from the down-sampling path for which the model learns to focus on target regions of varying shapes and sizes. Moreover, the residual connections help to train deeper models by ensuring better gradient flow. We conducted experiments on three datasets: CVC Clinic-DB, ISIC 2018, and the 2018 Data Science Bowl datasets and achieved Dice Coefficient scores of \textbf{94.35\%}, \textbf{91.68\%} and \textbf{92.45\%} respectively. Our results suggest that AttResDU-Net can be facilitated as a reliable method for automatic medical image segmentation in practice.

\end{abstract}


\section{Introduction}\label{sec:introduction}
Medical image segmentation is a crucial task for the automation of medical image-based diagnosis, as it enables the identification of diagnostically important anatomical structures, such as the heart, kidney cells, skin lesions, and polyps, etc. \cite{lei}. Segmentation of these structures is necessary for several medical tasks, including classification and detection. Manual inspection of these structures is time-consuming, laborious, and requires complex clinical experience due to the various morphological features and sizes. To overcome these limitations, computer-aided approaches that automatically segment the Region of Interest (RoI) could be helpful, as they can pick up on slight pixel variations that may not be perceivable to the human eye \cite{multi-scale, pranet}. Furthermore, they can process large volumes of data quickly and accurately, allowing medical professionals to make accurate diagnosis and treatment plans. At the same time, these systems can reduce their workload, allowing them to focus on more complex cases that require further clinical expertise \cite{wang2022}. In this regard, the development of efficient and accurate automatic medical image segmentation methods is crucial for facilitating better diagnosis and improving patient outcomes.

With the advent of deep learning, the use of Convolutional Neural Networks (CNNs) for biomedical image segmentation became highly popular \cite{lei, wang2022}. Among the most successful approaches to medical image segmentation are U-Net \cite{unet}, Attention U-Net \cite{att_unet}, and Double U-Net \cite{dunet}. One of the major problems in medical image segmentation is the difficulty of achieving promising results on target regions with varying shapes, structures, and boundaries using earlier deep neural networks such as U-Net and Attention U-Net. This is due to their limited training capacity and the presence of fewer parameters \cite{dunet}. To address this limitation, larger models such as Double U-Net and Transformers have been introduced \cite{dunet, DS-TransUNet}, but they suffer from vanishing gradients during backpropagation \cite{resnet}. The problem is further worsened by the fact that these huge networks need to learn to segment irregular shapes from diverse locations taken under varying lighting conditions from only a limited number of medical image samples. Recent literature tend to address this scarcity of samples via various preprocessing and data augmentation techniques \cite{polar, 9810234}.

In order to address the aforementioned issues, 
we introduce Attention-based Residual Double U-Net (AttResDU-Net), a unique combination of architectural modifications inspired by the Double U-Net model \cite{dunet}, attention mechanism \cite{att_unet}, and residual connections \cite{resnet} trained on images pre-processed using Color Constancy \cite{cc}. Our main contributions can be summarized as follows:
    \begin{itemize}
        \item We utilize \textbf{Color Constancy (CC)} to normalize the diversity of lighting conditions in the medical image samples. This allows the model to achieve state-of-the-art (SOTA) even with fewer data augmentations over the standalone Double U-Net architecture.
        \item We incorporate \textbf{Attention Gates} that allow our model to retain more relevant spatial information by suppressing irrelevant feature representation from the down-sampling path of the encoding network resulting in lower computational cost and less focus on noise and artifacts.
        \item We integrate \textbf{Residual Connections} in between our encoder and decoder which help to train deeper models that enables us to identify complex shapes and structures while also ensuring better gradient flow and faster convergence.
    \end{itemize}

\section{Related Works}\label{sec:lit_review}
In the early days, medical image segmentation was carried out using traditional machine-learning methods. Celebi et al. applied unsupervised methods to segment out skin lesions using clustering algorithms \cite{Celebi}. On the other hand, Wong et al. implemented a stochastic region merging approach on a pixel and a regional level to extract the lesions from macroscopic images \cite{wong}. For automatic polyp segmentation, Gross et al. used a template matching algorithm using multi-scale filtering for edge detection, which was then compared to a set of elliptic templates \cite{gross}. These methods relied on extracting hand-crafted features that limit them to segmenting medical images on small datasets and hence fail to generalize on larger ones. With technological advancement and rapid increase in computational resources, various deep learning models have emerged for analyzing medical images.\

CNNs have recently performed excellently across several medical segmentation benchmarks \cite{litjens}. One of the most popular neural network architectures for biomedical image segmentation is U-Net \cite{unet}. It uses a series of convolutional operations in the encoding path for spatial information and similar operations in the decoding path to retain contextual information. Furthermore, it has skip connections between the encoder and decoder blocks so that spatial information can propagate deep into the network. This allows the network to learn context and precise localization simultaneously. Various modifications and improvements to the U-Net have been proposed consequently. Zhou et al. introduced U-Net++, where the encoder and decoder are connected via convolutional networks instead of simple concatenation \cite{unet++}. Oktay et al. proposed Attention U-Net, where they apply attention gates (AG) on the skip connections that allow the model to automatically learn how to focus more on target regions of various shapes and structures rather than having an equal focus on the entire image \cite{att_unet}. Models trained with AGs suppress irrelevant areas of the image and focus more on extracting essential features required for the specific task.\ 

The traditional U-Net has a straightforward structure of a decoding path followed by an encoding path. In the decoding path, the activation cannot be adequately up-sampled as spatial information gets lost. To overcome this, skip connections are added from the encoder blocks to the decoder blocks. But due to poor representation of features from the encoding path, irrelevant features are also concatenated in the skip connections. U-Net++ and Attention U-Net mitigate this problem by inserting convolution blocks and attention gates on the skip connections. However, these networks are relatively shallower, with fewer parameters for which they fail to extract more complex features. To address this issue, Jha et al. proposed a Double U-Net based on two U-Nets stacked on top of each other with some unique components such as a pre-trained VGG-19 model as the first encoder, a squeeze and excite block (SE), and an Atrous Spatial Pyramid Pooling(ASPP) \cite{aspp} block to capture more complex regions of interest \cite{dunet}. The concatenation of the outputs of the two U-Nets allows the model to train more effectively. It uses ASPP layers for contextual information and Squeeze and Excite blocks which provide channel-wise attention to produce more refined segmentation maps. However, it still fails to filter out the irrelevant features being concatenated in the skip connections. Thus, the Double U-Net leaves scope for spatial attention to be implemented and filters out those irrelevant features through skip connections. Furthermore, it is a very large network with a significant number of parameters. To ensure better gradient flow and quick convergence, residual connections in the encoding and decoding blocks can be considered.\

More recent image segmentation architectures include transformers and multi-scale fusion networks. Chen et al. incorporate both U-Net and Transformers that solves the problem of U-Net alone not being able to capture long-range dependencies as convolution operations are generally suitable for modeling local information only \cite{trans, DS-TransUNet}. While the performance of these methods is impressive, the sheer amount of parameters in transformers and increased dataset size using augmentations make it computationally expensive. Srivastava et al. introduced a residual fusion network that is relatively small in size, having the ability to exchange multi-scale features that ensure better propagation of high and low-level features \cite{msrf}. However, as mentioned by the authors, the model fails to perform well on very low-contrast images. Hence, the use of preprocessing techniques is required to make the images consistent.

The aforementioned discussion necessitates the realization of a medical image segmentation architecture that utilizes attention gates to focus on relevant areas of the image to extract essential features, while also identifying irregular shapes via larger and deeper networks avoiding vanishing gradient problem via skip connections.

\section{Methodology}\label{sec:methodology}
\subsection{Overview of our proposed architecture}
\begin{figure*}[t]
\centering
\includegraphics[width = \textwidth]{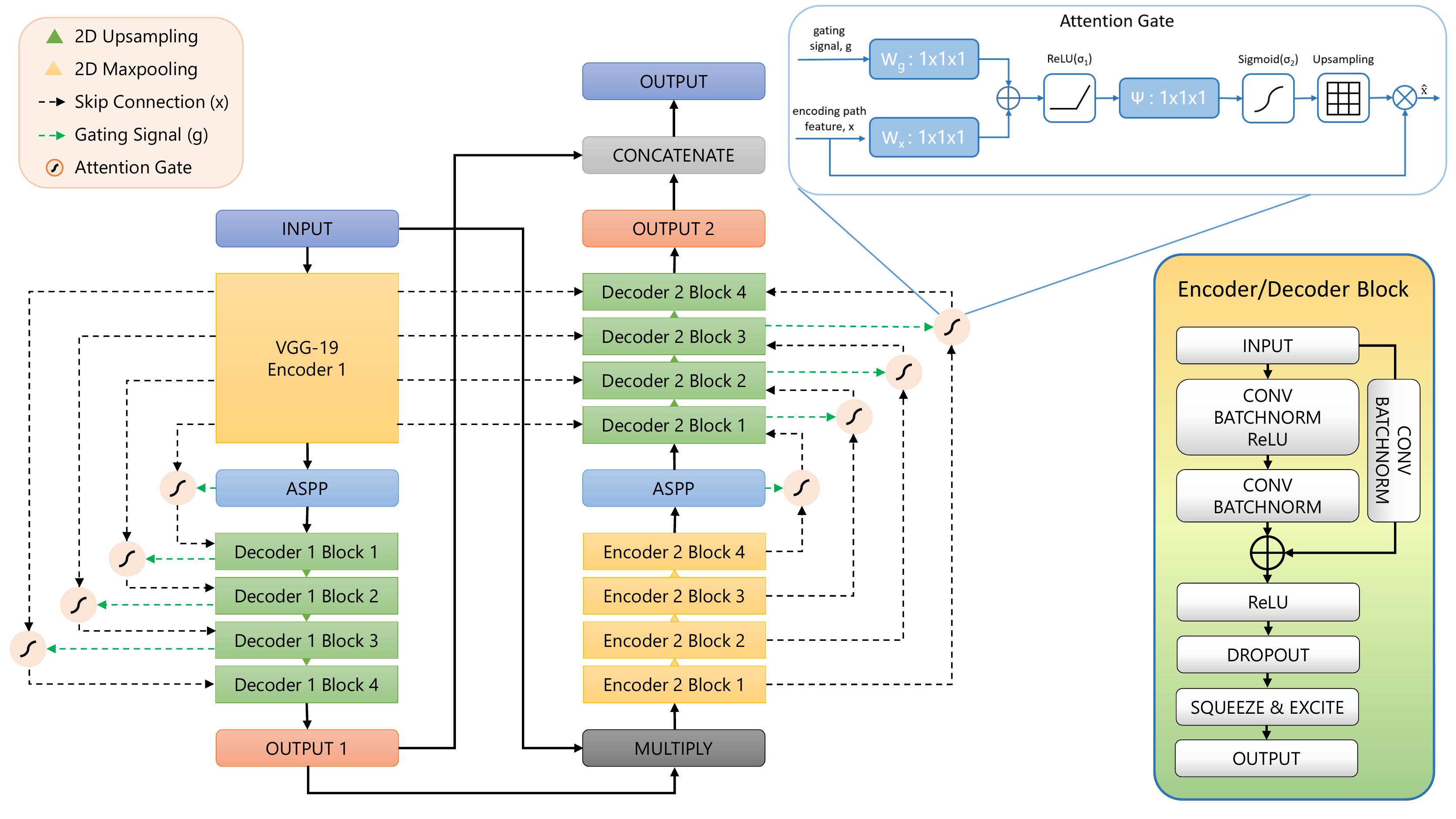}
\caption[Block diagram of the proposed AttResDU-Net]{Block diagram of the proposed AttResDU-Net}
\label{fig:our}
\end{figure*}

Figure \ref{fig:our} illustrates the block diagram for our proposed architecture. Initially, we apply Color Constancy as a preprocessing technique to the input images before feeding them to our network. The input image is then fed into Encoder 1, a pre-trained VGG-19\cite{vgg19} network, followed by an ASPP block. The ASPP block helps retain contextual information by re-sampling activation maps at different rates before applying convolutions\cite{aspp}. The activation maps are then passed to Decoder 1, a series of blocks that perform upsampling operations to generate the first output. This output is multiplied by the input image to help mitigate the loss of spatial information that might occur through the downsampling path. It then passes through another Encoder-ASPP-Decoder path to produce a second output. Finally, the first and second outputs are concatenated and passed through a convolutional block to generate the final segmentation map. We place attention-based skip connections between Encoder1 - Decoder1 and Encoder2 - Decoder2 to retain more relevant spatial information. Each encoder/decoder block comprises Squeeze and Excite blocks\cite{se}, which provide channel-wise attention that filters out which channel information is more relevant.

\subsection{Data preprocessing: Color Constancy}

\begin{figure}[htb]
    \centering
    \scalebox{0.8}{
      \includegraphics[width = 0.45\textwidth]{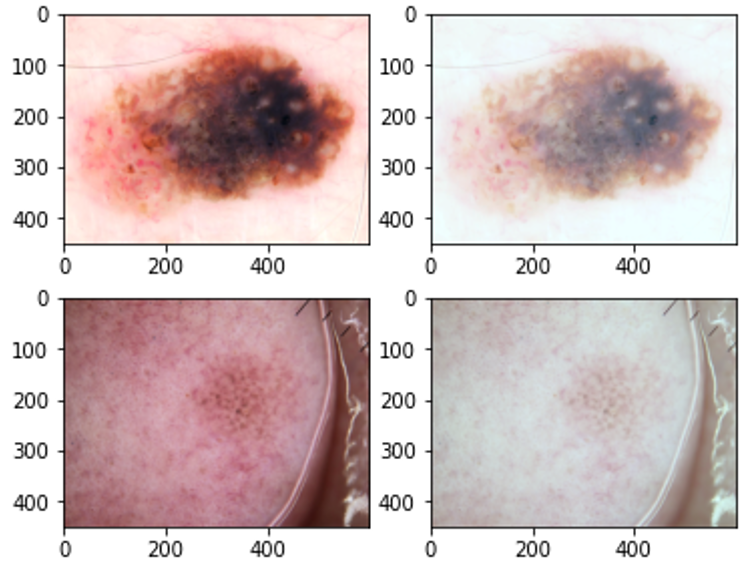}}
      \caption[Effect of applying shades of gray algorithm]{Example image in inconsistent color and lighting condition (left) normalized with Color Constancy (right)}
      \label{fig:cc}
\end{figure}

Dermoscopy and colonoscopy images are captured on different camera devices and under different light sources, resulting in variable illumination conditions. These conditions impact the generalization ability of the model as it may focus on learning color variations which do not play any role in segmentation tasks. There are also places of high reflectance in the images containing polyp, introducing redundant noise and other artifacts. These issues can be alleviated if we can normalize the diversity of lighting conditions in the dataset using the Color Constancy (CC) algorithm. Existing work, such as \cite{cc} has also shown that segmentation models perform better with CC as a preprocessing step. There are several algorithms for CC, however, we utilized the shades of gray algorithm \cite{sog}, one of the most popular preprocessing techniques used in the lesion segmentation task of the ISIC 2018 Challenge \cite{codella}. This approach is formed from the notion that CC would perform better if the scene average, i.e. the color to which different images are brought is a shade of gray. The formulation is shown in \equationautorefname~\ref{eq:sog}.
    
    \begin{align}
        \left(\frac{\int_{}^{}f(x)^p dx}{\int_{}^{} dx}\right) = K_e\label{eq:sog}
    \end{align}

Here $p$ is the Minkowski Norm, with shades of gray working best with $p = 6$ \cite{cc}. Figure \ref{fig:cc} shows the effect of applying the shades of gray algorithm on a skin lesion image.

 \begin{figure*}[h]
    \centering
    \begin{subfigure}{0.48\textwidth}
        \centering
        \includegraphics[width=\textwidth]{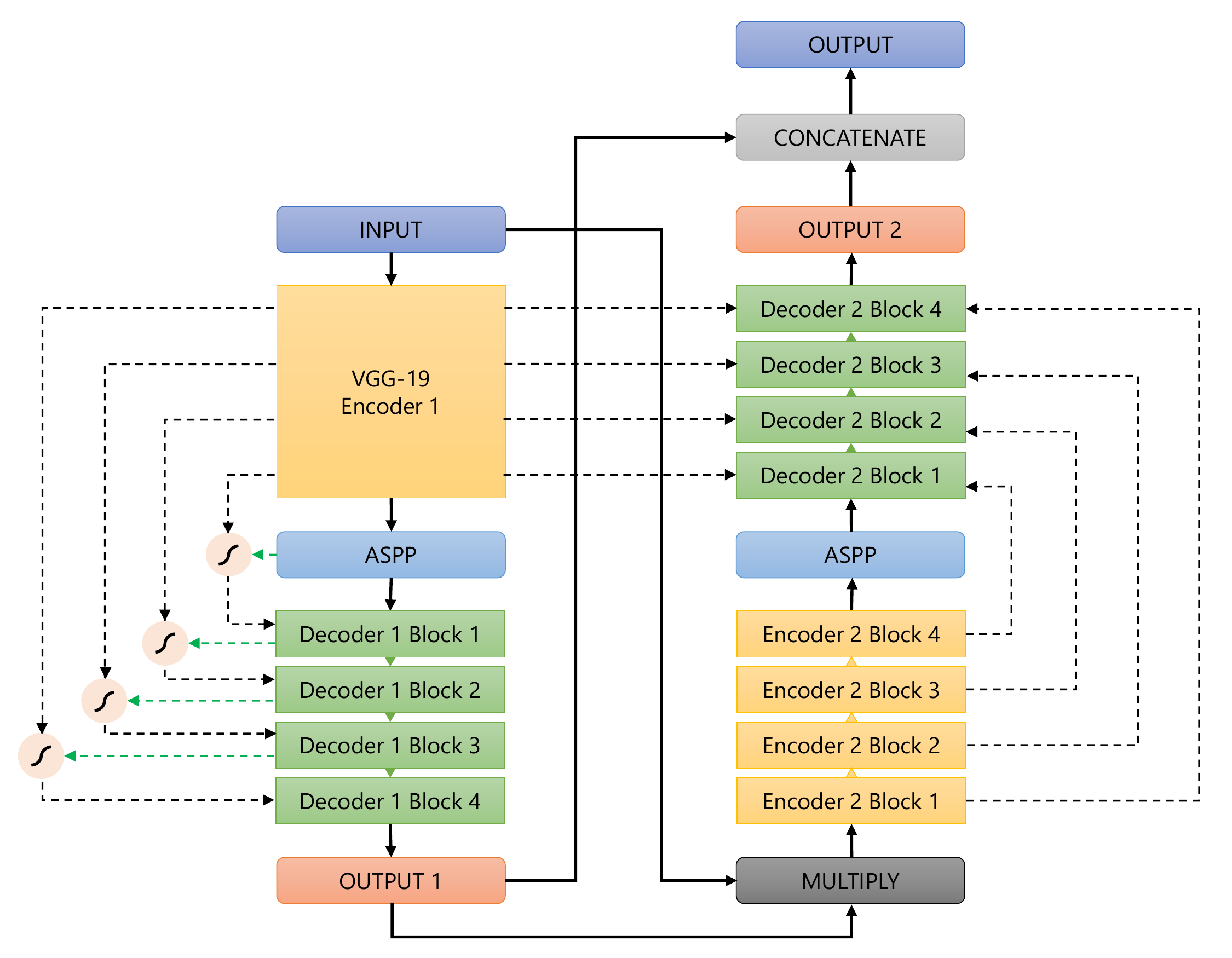}
        \caption[Block diagram of the Half-Attention Double U-Net]{Block diagram of the Half-Attention Double U-Net}
        \label{fig:ha}
    \end{subfigure}
    \begin{subfigure}{0.48\textwidth}
        \centering
        \includegraphics[width=\textwidth]{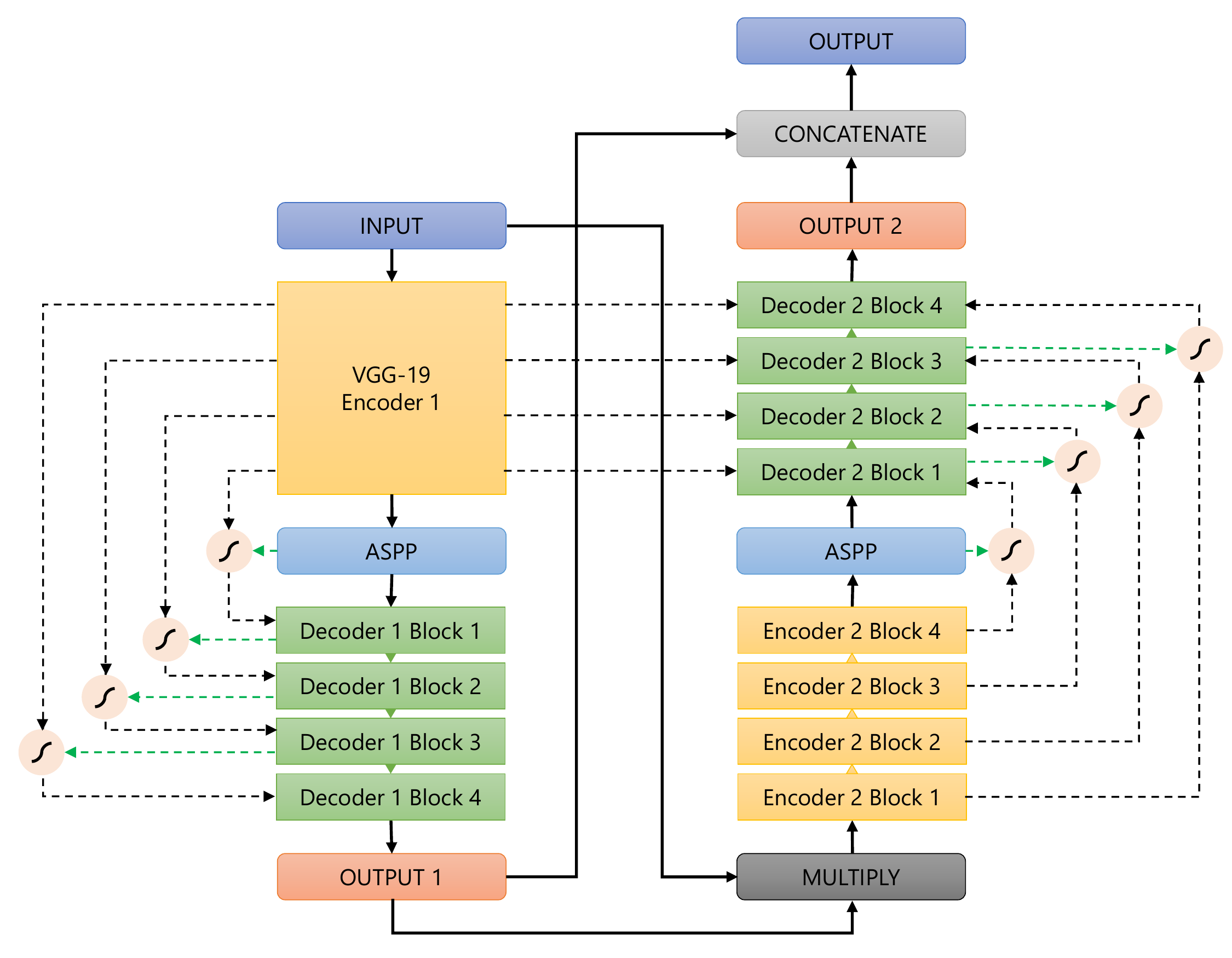}
        \caption[Block diagram of the Full-Attention Double U-Net]{Block diagram of the Full-Attention Double U-Net}
        \label{fig:fa}
    \end{subfigure}
    \caption[Network variants of our proposed architecture]{Network variants of our proposed architecture}
\end{figure*}

\subsection{Attention Gates}
We incorporate spatial attention to suppress the irrelevant features coming from the down-sampling path in the skip connections. Attention gates (AGs) adapted from \cite{att_unet} help the model focus on salient activations by prioritizing relevant regions. This can help focus computation on the most relevant part of the image, allowing the model to perform well while using fewer computational resources. At the same time, considering the complex structure of the segmentation masks that are required to be produced by the model, AGs can provide useful insights to identify the relevant information from the activation map. 

With a view to extracting such relevant information, we place the AGs in between the encoder and decoder in the skip connection path. The gates generate a weight matrix which is multiplied element-wise with the input coming from the encoder block. We explore two network variants of AGs: Half-Attention (\figureautorefname~\ref{fig:ha}), where we add AGs between Encoder 1-Decoder 1 only, and Full-Attention (\figureautorefname~\ref{fig:fa}), where add AGs between both Encoder 1-Decoder 1 and Encoder 2-Decoder 2.

\subsection{Encoder/Decoder Block}
The Encoder/Decoder Block in \figureautorefname~\ref{fig:our} illustrates the components of the convolution blocks in our proposed model. Here, the input passes through two separate paths. The first path is a series of CONV-BATCHNORM-ReLU-CONV-BATCHNORM operations sequentially to produce an intermediate output. As deeper networks fail to propagate small changes in derivatives to earlier layers, they suffer from the vanishing gradient problem \cite{resnet}. To help facilitate proper gradient flow and smoother convergence of the network, we add residual connections in the second path, which is a parallel path where the initial input goes through a single CONV-BATCHNORM operation before being added to the intermediate output produced from the first path. The concatenated output is then passed through the subsequent blocks. The output of each encoder block is downsampled through a Max-Pooling Layer, while the output of each decoder block is upsampled through an upsampling layer. 


\subsection{Loss Function}
To train and evaluate our model on each of the datasets we used Dice loss (\equationautorefname~\ref{eq:dice}), a region-based loss function. Dice loss helps with the class imbalance problem when the foreground class is much smaller compared to the background class. This is very common in medical images, as the RoI covers only a small area of the scan. 
\begin{align}
    Dice_{loss} = 1 - \frac{2 y \hat{y} + \lambda}{y + \hat{y} + \lambda}\label{eq:dice}
\end{align}

Here, $y$ is the ground truth value, and $\hat{y}$ is the predicted value.

\section{Experimental Analysis}\label{sec:results}

\subsection{Dataset}
There are several datasets in the medical image domain that are regarded as benchmark datasets. Among these datasets, we used the ISIC2018 (Skin Lesion), CVC-ClinicDB (Polyp), and 2018 Data Science Bowl (Nuclei) datasets for training, validating, and testing our proposed methods and architectures. The details of the datasets are as follows:
        
        \begin{itemize}
            \item CVC-ClinicDB \cite{CVCClinicDB1} is a database containing colonoscopy images and their corresponding manually annotated binary segmentation masks which represent the region of the image covered by polyps. The dataset contains 612 colonoscopy images of size 388x288 pixels.
            
            \item The ISIC 2018 Challenge \cite{isic2018} on Skin Lesion Analysis Towards Melanoma Detection was divided into three parts among which the first task was skin lesion segmentation. This dataset includes 2594 dermoscopic images along with their corresponding professionally annotated binary segmentation masks.
            
            \item The 2018 Data Science Bowl \cite{DataScienceBowl1} dataset contains a diverse collection of segmented nuclei images acquired under a variety of conditions and from a range of organisms. This dataset is designed to challenge a model's generalizability across these variations. It contains 670 nuclei images and the segmented masks of each nucleus. Each mask contains one nucleus and is not allowed to overlap, i.e. no pixel belongs to two different masks.
        \end{itemize}

\subsection{Experimental Setup}
We trained our model and performed all experiments on Kaggle using Tesla V100 GPU. All of the datasets were split into an 80-10-10 train-valid-test split. We used the original image size for the CVC-ClinicDB (polyp) and 2018 Data Science Bowl (nuclei) datasets since they contained a low number of images and did not require much time to train. However, for the ISIC 2018 (skin lesion) dataset, we resized the images to 192$\times$256 for faster training time, as this dataset contained images with inconsistent resolutions and required longer training times. We employed Dice Loss \cite{sudre2017generalised} as the loss function, which addresses the class imbalance problem in segmentation tasks, and we used the Nadam \cite{dozat} optimizer for each dataset. The learning rates were varied according to the size of the dataset, with 1e-4 for both the ISIC 2018 (skin lesion) and CVC-ClinicDB (polyp) dataset, while we used a lower learning rate of 1e-5 for the 2018 Data Science Bowl (nuclei) dataset since the model would run out of data before convergence. All models are trained for 40 epochs with early stopping to prevent overfitting and learning rate schedules for reducing the learning rate when the model stops improving.\

Alongside applying CC, we performed normalization and sample-wise centering. Random rotation, vertical and horizontal flip, Hue Saturation Value (HSV) conversion, random brightness-contrast, and histogram equalization were among the data augmentation techniques we applied to each of the datasets to increase the dataset sizes by six-fold using the augmentation tool: Albumentations \cite{album}. The code for our proposed model can be found at \href{https://github.com/fkhan98/Medical-Image-Segmentation-with-Attention-based-Residual-Double-U-Net}{https://github.com/fkhan98/Medical-Image-Segmentation-with-Attention-based-Residual-Double-U-Net}.

\subsection{Evaluation Metrics}
The measure of the Dice Coefficient (DSC) \cite{zou2004statistical} is the standard evaluation metric for segmentation. It represents the similarity between the generated RoIs and those of the ground truth. Since these RoIs are significantly smaller than the background class, a class imbalance problem arises. Using the dice coefficient as the evaluation metric helps to alleviate the sensitivity to this class imbalance. It gives a higher value when more of the generated RoI pixels match the ground-truth RoI pixels, irrespective of the background class.

Apart from this, we have used other evaluation metrics, including Intersection over Union (IOU), a region-based metric similar to DSC, and Precision and Recall, which consider the overlooked cases and false alarms. These four metrics are widely used in image segmentation literature and are, therefore, suitable for our experiments.

\begin{align}
    DSC &= \frac{2*TP}{(TP+FP)+(TP+FN)}\\
    IOU &= \frac{TP}{TP + FP + FN}\\
    Recall &= \frac{TP}{TP+FN}\\
    Precision &= \frac{TP}{TP+FP}
\end{align}

Here, TP = True Positive, FP = False Positive, FN = False Negative

\subsection{Ablation Study}
We conducted ablation studies incorporating three different modules: Color Constancy (CC) preprocessing, Attention Gates (AGs), and Residual Connections, to determine which overall pipeline works best. We divided the ablation study into two steps. In the first step, we examined which of our model variants performs better. Based on the outcome of the first step, we conducted experiments with that variant to determine the effect of CC preprocessing and residual connections to determine the best overall pipeline.

\begin{table}[htb]
\centering
\caption{Ablation Experiments on ISIC2018 dataset (\%)}
\label{tab:ablation}
\begin{tabular}{c|cccc}
\hline
\textbf{Method} & \textbf{DSC} & \textbf{IOU} & \textbf{Recall} & \textbf{Precision} \\ \hline
Half-Attention  & 90.90            & 83.38            & 85.79            & 94.63  \\
Full-Attention  & \textbf{91.64} & \textbf{84.63}     & \textbf{86.13}   & \textbf{95.76}  \\ \hline
AttDU-Net       & 90.73          & 83.14              & \textbf{88.66}   & 92.54  \\
AttDU-Net + CC    & 91.64          & 84.63            & 86.13            & \textbf{95.76} \\
AttResDU-Net + CC & \textbf{91.68} & \textbf{84.68}   & 87.55            & 94.19 \\ \hline
\end{tabular}
\end{table}

\textbf{Effect of AGs in Encoding-Decoding Paths:}
The first set of ablation experiments compares the two variants of our architecture: Half-Attention and Full-Attention. Table \ref{tab:ablation} indicates that the Full-Attention variant results in better performance (DSC \textbf{91.64\%}, IOU \textbf{84.63\%}, Recall \textbf{86.13\%}, and Precision \textbf{95.76\%}) compared to its counterpart (DSC 90.9\%, IOU 83.38\%, Recall 85.79\%, and Precision 94.63\%). Adding more attention gates in the skip connections allows for more relevant spatial information to be harnessed, providing better segmentation results. Regarding computational complexity, Half-Attention has 35 million parameters with a Floating Point Operations Per Second (FLOPS) count of 90.1 GFLOPS. At the same time, the Full-Attention stands at 92.1 GFLOPS with a parameter count of 36.5 million.

\textbf{Effect of CC and Residual Connections:}
Based on the outcome of the previous experiments, we conducted the second set of ablation experiments using the Full-Attention variant of our proposed model. As shown in Table \ref{tab:ablation}, we observe that applying CC improves our model's DSC by \textbf{0.91\%}, IOU by \textbf{1.49\%}, and Precision by \textbf{3.22\%}. This shows that transforming images to appear under a uniform light source helps the model achieve better segmentation results. Moreover, adding residual connections slightly enhances the DSC to \textbf{91.68\%} and IOU to \textbf{84.68\%} while stabilizing the training process, resulting in faster convergence to the optima. The model containing residual connections converged faster (\textbf{20 epochs}) than the one without residual connections (30 epochs). This leads us to conclude that the Full-Attention Double U-Net model with residual connections across the convolution blocks trained on images preprocessed by applying CC generates the best possible results.

\begin{table*}[htb]
\addtolength{\tabcolsep}{-2.25pt}
\center
\caption{Performance of our model against the SOTA on three benchmark datasets: CVC-ClinicDB (polyp), ISIC-2018 (lesion), and 2018 Data Science Bowl (nuclei). The table, adapted from \cite{msrf}, denotes the estimated mean score (\%) of the respective models}
\scalebox{1.2}{
\begin{tabular}{ c|cccc|cccc|cccc|c}

 &
  \multicolumn{4}{c|}{\textbf{CVC-ClinicDB}} &
  \multicolumn{4}{c|}{\textbf{ISIC-2018}} &
  \multicolumn{4}{c|}{\textbf{2018 Data-Science Bowl}}&
  \textbf{} \\ \hline
\textbf{Model} &
  \textbf{DSC} &
  \textbf{IOU} &
  \textbf{Rec} &
  \textbf{Prec} &
  \textbf{DSC} &
  \textbf{IOU} &
  \textbf{Rec} &
  \textbf{Prec} &
  \textbf{DSC} &
  \textbf{IOU} &
  \textbf{Rec} &
  \textbf{Prec} &
  \textbf{Params (M)} \\ \hline
U-Net \cite{unet} &
  87.81 &
  78.81 &
  78.65 &
  93.29 &
  85.54 &
  78.47 &
  82.04 &
  \textbf{94.74} &
  90.80 &
  83.14 &
  90.29 &
  91.30 &
  7.11\\ 
U-Net++ \cite{unet++} &
  84.53 &
  75.59 &
  89.17 &
  83.23 &
  80.94 &
  72.88 &
  78.66 &
  90.84 &
  77.05 &
  52.65 &
  71.59 &
  66.57 &
  9.04\\ 
DeepLabv3+ \cite{deeplab} &
  88.97 &
  87.06 &
  92.51 &
  93.66 &
  87.72 &
  81.28 &
  86.81 &
  92.72 &
  88.57 &
  83.67 &
  91.41 &
  90.81 &
  41.25\\ 
\begin{tabular}[c]{@{}c@{}}Res \\ U-Net++\end{tabular} \cite{res-unet++} &
  90.75 &
  85.87 &
  91.56 &
  93.25 &
  85.57 &
  81.35 &
  88.01 &
  86.76 &
  90.98 &
  83.70 &
  91.69 &
  90.57 &
  4.07\\ 
Pra-Net \cite{pranet}&
  90.72 &
  85.75 &
  92.27 &
  91.34 &
  - &
  - &
  - &
  - &
  87.51 &
  78.68 &
  91.82 &
  84.38 &
  32.54\\ 
UACA-Net \cite{uacanet} &
  90.98 &
  86.49 &
  91.74 &
  91.14 &
  - &
  - &
  - &
  - &
  86.88 &
  77.91 &
  90.61 &
  84.14 &
  69.15\\ 
\begin{tabular}[c]{@{}c@{}}Double\\ U-Net\end{tabular} \cite{dunet} &
  92.39 &
  86.11 &
  84.57 &
  95.92 &
  89.62 &
  82.12 &
  87.80 &
  94.59 &
  91.33 &
  84.07 &
  64.07 &
  94.96 &
  29.29\\ 
\begin{tabular}[c]{@{}c@{}}DS-Trans\\ U-Net\end{tabular} \cite{DS-TransUNet} &
  94.22 &
  89.39 &
  95.00 &
  93.69 &
  91.32 &
  85.23 &
  92.17 &
  92.71 &
  92.19 &
  \textbf{86.12} &
  93.78 &
  91.24 &
  (88$\sim$197)\\ 
MSRF-Net \cite{msrf}&
  94.20 &
  \textbf{90.43} &
  \textbf{95.67} &
  94.27 &
  88.24 &
  83.73 &
  88.93 &
  93.48 &
  92.24 &
  85.34 &
  \textbf{94.02} &
  90.22 &
  18.38\\ 
\begin{tabular}[c]{@{}c@{}}Polar Res\\ U-Net++\end{tabular} \cite{polar}&
  93.74 &
  89.77 &
  93.68 &
  94.88 &
  \textbf{92.53} &
  \textbf{87.43} &
  \textbf{94.64} &
  92.53 &
  - &
  - &
  - &
  - &
  -\\ 
\begin{tabular}[c]{@{}c@{}}Full Att-Res\\ DU-Net (Ours)\end{tabular} &
  \textbf{94.35} &
  89.32 &
  87.50 &
  \textbf{97.37} &
  91.68 &
  84.68 &
  87.55 &
  94.19 &
  \textbf{92.45} &
  85.96 &
  65.55 &
  \textbf{96.29} &
  36.50\\ \hline
\end{tabular}}
\label{table1}
\end{table*}

\subsection{Comparison with state-of-the-art methods}
We evaluated our proposed architecture on three different datasets and compared our results with state-of-the-art models as shown in Table \ref{table1}. In the CVC-ClinicDB dataset, our model outperforms (DSC \textbf{94.35\%}) the best performing DS-Trans-U-Net (DSC \textbf{94.22\%}) along with a significant improvement in precision (\textbf{3.93\%} increase). Moreover, DS-Trans-U-Net is built upon the swin transformer with a vast number of parameters ranging from \textbf{88$\sim$197} million depending on the variant, making it computationally expensive to train \cite{swin} while our model stands at 36.5 million parameters. With a \textbf{1.96\%} improvement in DSC, our model also outperforms the Double U-Net, which requires a huge number of data augmentations (\textbf{50,000} training images) \cite{dunet}. We achieve comparable scores with MSRF-Net over all the metrics and perform better in low-contrast images, as seen in Figure \ref{qual2}, which is a limitation of their network \cite{msrf}. Our Full-Attention Res-DU-Net improves upon all the existing models discussed by using an attention mechanism and requires fewer images (\textbf{14,000}) to train with the help of CC.

To verify the robustness of our model, we further trained and tested it on two other datasets. Table \ref{table1} shows the results of our proposed model on the ISIC 2018 dataset compared to the current state-of-the-art results. Our Full Attention Res-DU-Net achieves comparable DSC, IOU, Precision, and Recall scores with all the existing methods. Although Polar Res-U-Net++ currently performs the best in this dataset, this method consists of a preprocessing technique that converts cartesian images to polar coordinate images using a center-point predictor network. Therefore, this method relies heavily on the preprocessing technique, which requires a \textbf{model-based preprocessing} approach both during training and inference \cite{polar}. In contrast, our method has much less overhead on preprocessing. Subsequently, our model achieves the best DSC score (\textbf{92.45\%}) and precision (\textbf{96.29\%}) on the 2018 Data Science Bowl dataset as well. From these experiments, we can conclude that our model generalizes successfully on segmentation tasks across various types of medical data.

\subsection{Qualitative Analysis}
\begin{figure}[t]
    \centering
    \begin{subfigure}{0.48\textwidth}
        \centering
        \includegraphics[width=\textwidth]{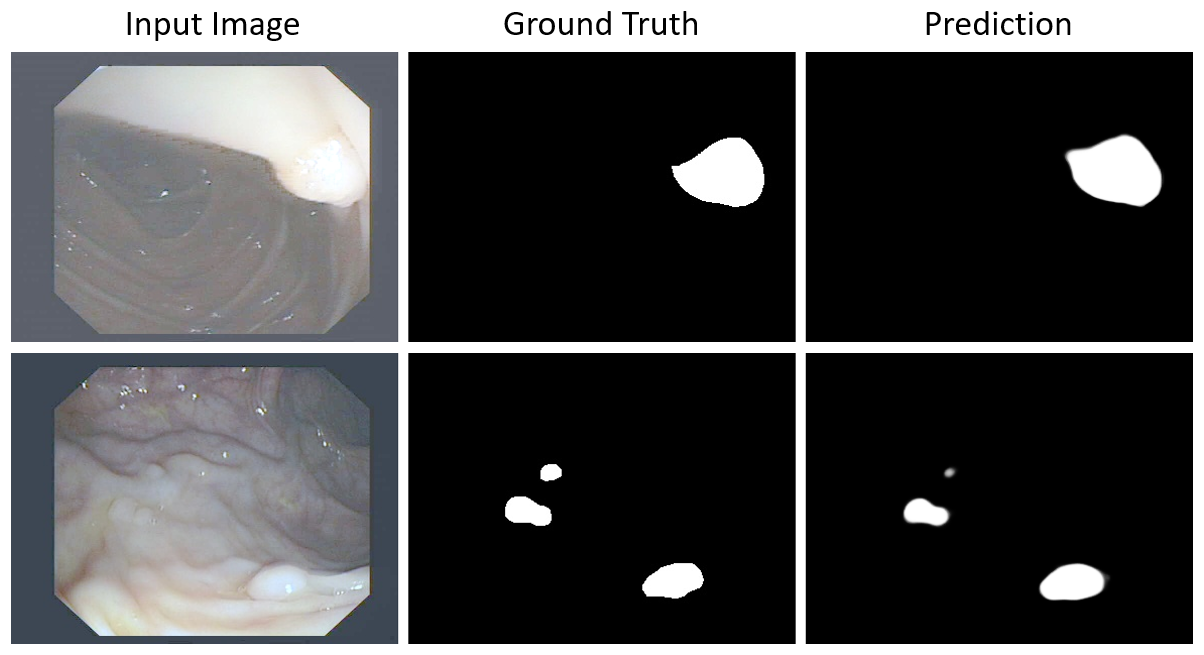}
        \caption{CVC-ClinicDB Polyp Dataset}
        \label{qual2}
    \end{subfigure}
    \begin{subfigure}{0.48\textwidth}
        \centering
        \includegraphics[width=\textwidth]{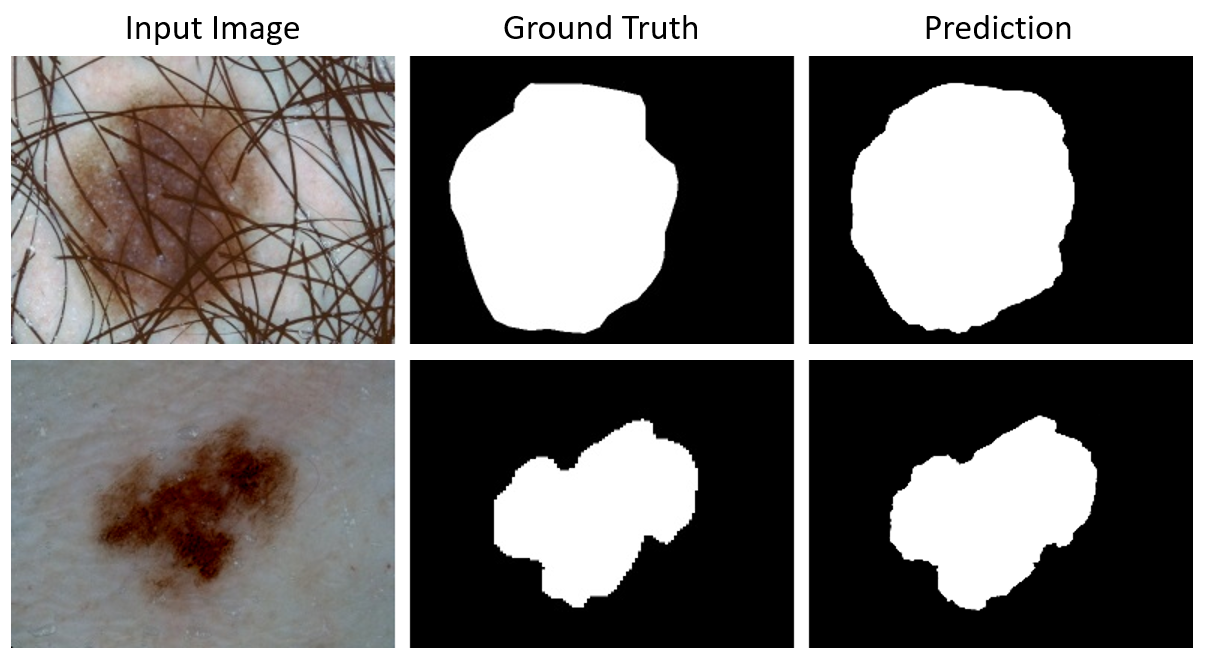}
        \caption{ISIC-2018 Skin Lesion Dataset}
        \label{qual1}
    \end{subfigure}
    \begin{subfigure}{0.48\textwidth}
        \centering
        \includegraphics[width=\textwidth]{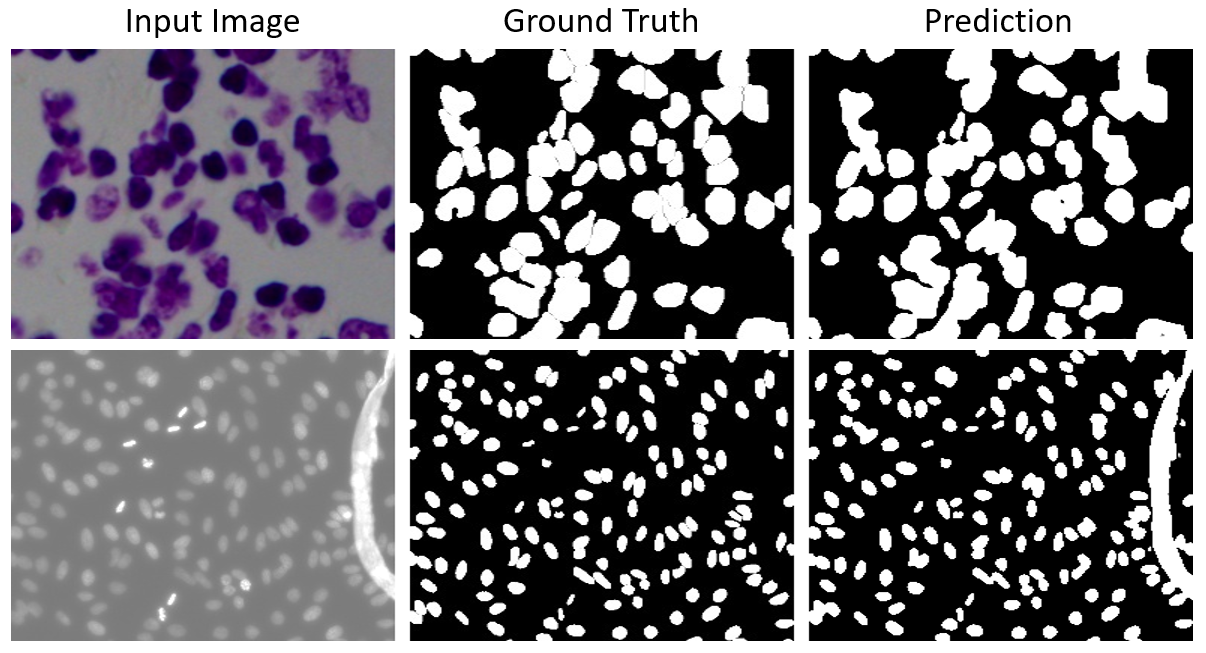}
        \caption{2018 Data Science Bowl Nuclei Dataset}
        \label{qual3}
    \end{subfigure}
    \caption[Qualitative results of our model]{Qualitative results of our model across three datasets}
    \label{fig:QualitativeResults}
\end{figure}

Figure \ref{fig:QualitativeResults} illustrates the outputs of our model for the three different datasets. We can observe that our model is able to segment out even flat polyp structures as shown in Figure \ref{qual2}. The application of CC helps to reduce the reflectance in each image and improves performance even on low contrast images. From Figure \ref{qual1} we can observe that irregular shape and size of skin lesions or hair artifacts are also not challenging as our model still produces high-quality segmentation masks for them. Moreover, our model is able to correctly segment out the densely connected nuclei cells as seen in the input image in \ref{qual3}, irrespective of the color of the image. From this qualitative analysis, we can conclude that our model mitigates the research challenges mentioned to a great extent.

\section{Conclusion and Future Work}\label{sec:conclusion}
In this paper, we propose an attention-based residual Double U-Net architecture for medical image segmentation. Our approach improves upon a Double U-Net architecture by adding attention gates to the skip connections, residual connections to the convolutional blocks, and the application of the Color Constancy algorithm to increase data efficiency. Thorough experiments show that our approach establishes state-of-the-art results in the CVC-ClinicDB and 2018 Data Science Bowl datasets while having scores similar to SOTA in the ISIC 2018 dataset. For further improvements, a simplified architecture can be designed to reduce the number of parameters to train the model faster while retaining similar accuracy. Integrating various boundary-aware loss functions to accurately predict non-uniform boundaries can also be an avenue to look towards.


\bibliographystyle{IEEEtran}
\bibliography{References}

\begin{thebibliography}{10}
\providecommand{\url}[1]{#1}
\csname url@samestyle\endcsname
\providecommand{\newblock}{\relax}
\providecommand{\bibinfo}[2]{#2}
\providecommand{\BIBentrySTDinterwordspacing}{\spaceskip=0pt\relax}
\providecommand{\BIBentryALTinterwordstretchfactor}{4}
\providecommand{\BIBentryALTinterwordspacing}{\spaceskip=\fontdimen2\font plus
\BIBentryALTinterwordstretchfactor\fontdimen3\font minus
  \fontdimen4\font\relax}
\providecommand{\BIBforeignlanguage}[2]{{%
\expandafter\ifx\csname l@#1\endcsname\relax
\typeout{** WARNING: IEEEtran.bst: No hyphenation pattern has been}%
\typeout{** loaded for the language `#1'. Using the pattern for}%
\typeout{** the default language instead.}%
\else
\language=\csname l@#1\endcsname
\fi
#2}}
\providecommand{\BIBdecl}{\relax}
\BIBdecl

\bibitem{lei}
\BIBentryALTinterwordspacing
R.~Wang, T.~Lei, R.~Cui, B.~Zhang, H.~Meng, and A.~K. Nandi, ``{Medical image
  segmentation using deep learning: A survey},'' \emph{{IET Image Processing}},
  vol.~16, no.~5, pp. 1243--1267, 2022. [Online]. Available:
  \url{https://ietresearch.onlinelibrary.wiley.com/doi/abs/10.1049/ipr2.12419}
\BIBentrySTDinterwordspacing

\bibitem{multi-scale}
\BIBentryALTinterwordspacing
Z.~Wei, F.~Shi, H.~Song, W.~Ji, and G.~Han, ``{Attentive boundary aware network
  for multi-scale skin lesion segmentation with adversarial training},''
  \emph{{Multimedia Tools and Applications}}, vol.~79, no.~37, pp.
  27\,115--27\,136, 2020. [Online]. Available:
  \url{https://link.springer.com/article/10.1007/s11042-020-09334-2}
\BIBentrySTDinterwordspacing

\bibitem{pranet}
\BIBentryALTinterwordspacing
D.-P. Fan, G.-P. Ji, T.~Zhou, G.~Chen, H.~Fu, J.~Shen, and L.~Shao, ``{Pranet:
  Parallel reverse attention network for polyp segmentation},'' in
  \emph{International conference on medical image computing and
  computer-assisted intervention}.\hskip 1em plus 0.5em minus 0.4em\relax
  {Springer}, 2020, pp. 263--273. [Online]. Available:
  \url{https://arxiv.org/abs/2006.11392}
\BIBentrySTDinterwordspacing

\bibitem{wang2022}
\BIBentryALTinterwordspacing
R.~Wang, T.~Lei, R.~Cui, B.~Zhang, H.~Meng, and A.~K. Nandi, ``{Medical image
  segmentation using deep learning: A survey},'' \emph{IET Image Processing},
  vol.~16, no.~5, pp. 1243--1267, 2022. [Online]. Available:
  \url{https://ietresearch.onlinelibrary.wiley.com/doi/abs/10.1049/ipr2.12419}
\BIBentrySTDinterwordspacing

\bibitem{unet}
\BIBentryALTinterwordspacing
O.~Ronneberger, P.~Fischer, and T.~Brox, ``{U-net: Convolutional networks for
  biomedical image segmentation},'' in \emph{International Conference on
  Medical image computing and computer-assisted intervention}.\hskip 1em plus
  0.5em minus 0.4em\relax {Springer}, 2015, pp. 234--241. [Online]. Available:
  \url{https://link.springer.com/chapter/10.1007/978-3-319-24574-4_28}
\BIBentrySTDinterwordspacing

\bibitem{att_unet}
\BIBentryALTinterwordspacing
O.~Oktay, J.~Schlemper, L.~L. Folgoc, M.~C.~H. Lee, M.~P. Heinrich, K.~Misawa,
  K.~Mori, S.~G. McDonagh, N.~Y. Hammerla, B.~Kainz, B.~Glocker, and
  D.~Rueckert, ``{Attention U-Net: Learning Where to Look for the Pancreas},''
  \emph{{CoRR}}, vol. abs/1804.03999, 2018. [Online]. Available:
  \url{http://arxiv.org/abs/1804.03999}
\BIBentrySTDinterwordspacing

\bibitem{dunet}
\BIBentryALTinterwordspacing
D.~Jha, M.~A. Riegler, D.~Johansen, P.~Halvorsen, and H.~D. Johansen,
  ``{DoubleU-Net: A Deep Convolutional Neural Network for Medical Image
  Segmentation},'' in \emph{2020 IEEE 33rd International Symposium on
  Computer-Based Medical Systems (CBMS)}.\hskip 1em plus 0.5em minus
  0.4em\relax Los Alamitos, CA, USA: {IEEE Computer Society}, jul 2020, pp.
  558--564. [Online]. Available:
  \url{https://doi.ieeecomputersociety.org/10.1109/CBMS49503.2020.00111}
\BIBentrySTDinterwordspacing

\bibitem{DS-TransUNet}
\BIBentryALTinterwordspacing
A.~Lin, B.~Chen, J.~Xu, Z.~Zhang, G.~Lu, and D.~Zhang, ``{DS-TransUNet: Dual
  Swin Transformer U-Net for Medical Image Segmentation},'' \emph{{IEEE Trans.
  Instrum. Meas.}}, vol.~71, pp. 1--15, 2022. [Online]. Available:
  \url{https://ieeexplore.ieee.org/document/9785614/}
\BIBentrySTDinterwordspacing

\bibitem{resnet}
\BIBentryALTinterwordspacing
K.~He, X.~Zhang, S.~Ren, and J.~Sun, ``{Deep Residual Learning for Image
  Recognition},'' in \emph{2016 {IEEE} Conference on Computer Vision and
  Pattern Recognition, {CVPR} 2016, Las Vegas, NV, USA, June 27-30,
  2016}.\hskip 1em plus 0.5em minus 0.4em\relax {IEEE Computer Society}, 2016,
  pp. 770--778. [Online]. Available:
  \url{https://ieeexplore.ieee.org/document/7780459/}
\BIBentrySTDinterwordspacing

\bibitem{polar}
\BIBentryALTinterwordspacing
M.~Bencevic, I.~Galic, M.~Habijan, and D.~Babin, ``{Training on Polar Image
  Transformations Improves Biomedical Image Segmentation},'' \emph{{IEEE
  Access}}, vol.~9, pp. 133\,365--133\,375, 2021. [Online]. Available:
  \url{https://doi.org/10.1109/ACCESS.2021.3116265}
\BIBentrySTDinterwordspacing

\bibitem{9810234}
S.~Ahmed, M.~B. Hasan, T.~Ahmed, M.~R.~K. Sony, and M.~H. Kabir, ``{Less is
  More: Lighter and Faster Deep Neural Architecture for Tomato Leaf Disease
  Classification},'' \emph{{IEEE Access}}, vol.~10, pp. 68\,868--68\,884, 2022.

\bibitem{cc}
\BIBentryALTinterwordspacing
J.~Ng, M.~Goyal, B.~Hewitt, and M.~H. Yap, ``{The effect of color constancy
  algorithms on semantic segmentation of skin lesions},'' in \emph{Medical
  Imaging 2019: Biomedical Applications in Molecular, Structural, and
  Functional Imaging, San Diego, California, United States, 16-21 February
  2019}, ser. {SPIE} Proceedings, B.~Gimi and A.~Kr{\'{o}}l, Eds., vol.
  10953.\hskip 1em plus 0.5em minus 0.4em\relax {SPIE}, 2019, p. 109530R.
  [Online]. Available:
  \url{https://www.spiedigitallibrary.org/conference-proceedings-of-spie/10953/2512702/The-effect-of-color-constancy-algorithms-on-semantic-segmentation-of/10.1117/12.2512702.full?SSO=1}
\BIBentrySTDinterwordspacing

\bibitem{Celebi}
\BIBentryALTinterwordspacing
M.~E. Celebi, W.~Guo, Y.~A. Aslandogan, and P.~R. Bergstresser, ``{Skin Lesion
  Segmentation Using Clustering Techniques},'' in \emph{Proceedings of the
  Eighteenth International Florida Artificial Intelligence Research Society
  Conference, Clearwater Beach, Florida, {USA}}, I.~Russell and Z.~Markov,
  Eds.\hskip 1em plus 0.5em minus 0.4em\relax {AAAI Press}, 2005, pp. 364--369.
  [Online]. Available:
  \url{http://www.aaai.org/Library/FLAIRS/2005/flairs05-060.php}
\BIBentrySTDinterwordspacing

\bibitem{wong}
\BIBentryALTinterwordspacing
A.~Wong, J.~Scharcanski, and P.~W. Fieguth, ``{Automatic Skin Lesion
  Segmentation via Iterative Stochastic Region Merging},'' \emph{{IEEE Trans.
  Inf. Technol. Biomed.}}, vol.~15, no.~6, pp. 929--936, 2011. [Online].
  Available: \url{https://ieeexplore.ieee.org/document/5776681}
\BIBentrySTDinterwordspacing

\bibitem{gross}
\BIBentryALTinterwordspacing
S.~Gross, M.~Kennel, T.~Stehle, J.~Wulff, J.~J.~W. Tischendorf, C.~Trautwein,
  and T.~Aach, ``{Polyp Segmentation in NBI Colonoscopy},'' in
  \emph{Bildverarbeitung f{\"{u}}r die Medizin 2009: Algorithmen - Systeme -
  Anwendungen, Proceedings des Workshops vom 22. bis 25. M{\"{a}}rz 2009 in
  Heidelberg}, ser. Informatik Aktuell, H.~Meinzer, T.~M. Deserno, H.~Handels,
  and T.~Tolxdorff, Eds.\hskip 1em plus 0.5em minus 0.4em\relax {Springer},
  2009, pp. 252--256. [Online]. Available:
  \url{https://link.springer.com/chapter/10.1007/978-3-540-93860-6_51}
\BIBentrySTDinterwordspacing

\bibitem{litjens}
\BIBentryALTinterwordspacing
G.~Litjens, T.~Kooi, B.~E. Bejnordi, A.~A.~A. Setio, F.~Ciompi, M.~Ghafoorian,
  J.~A. W.~M. van~der Laak, B.~van Ginneken, and C.~I. S{\'{a}}nchez, ``{A
  survey on deep learning in medical image analysis},'' \emph{{Medical Image
  Analysis}}, vol.~42, pp. 60--88, 2017. [Online]. Available:
  \url{https://www.sciencedirect.com/science/article/abs/pii/S1361841517301135}
\BIBentrySTDinterwordspacing

\bibitem{unet++}
\BIBentryALTinterwordspacing
Z.~Zhou, M.~M.~R. Siddiquee, N.~Tajbakhsh, and J.~Liang, ``{UNet++: A Nested
  U-Net Architecture for Medical Image Segmentation},'' in \emph{Deep Learning
  in Medical Image Analysis - and - Multimodal Learning for Clinical Decision
  Support - 4th International Workshop, {DLMIA} 2018, and 8th International
  Workshop, {ML-CDS} 2018, Held in Conjunction with {MICCAI} 2018, Granada,
  Spain, September 20, 2018, Proceedings}, ser. Lecture Notes in Computer
  Science, D.~Stoyanov, Z.~Taylor, G.~Carneiro, T.~F. Syeda{-}Mahmood, A.~L.
  Martel, L.~Maier{-}Hein, J.~M. R.~S. Tavares, A.~P. Bradley, J.~P. Papa,
  V.~Belagiannis, J.~C. Nascimento, Z.~Lu, S.~Conjeti, M.~Moradi, H.~Greenspan,
  and A.~Madabhushi, Eds., vol. 11045.\hskip 1em plus 0.5em minus 0.4em\relax
  {Springer}, 2018, pp. 3--11. [Online]. Available:
  \url{https://link.springer.com/chapter/10.1007/978-3-030-00889-5_1}
\BIBentrySTDinterwordspacing

\bibitem{aspp}
\BIBentryALTinterwordspacing
L.~Chen, G.~Papandreou, F.~Schroff, and H.~Adam, ``{Rethinking Atrous
  Convolution for Semantic Image Segmentation},'' \emph{{CoRR}}, vol.
  abs/1706.05587, 2017. [Online]. Available:
  \url{http://arxiv.org/abs/1706.05587}
\BIBentrySTDinterwordspacing

\bibitem{trans}
\BIBentryALTinterwordspacing
J.~Chen, Y.~Lu, Q.~Yu, X.~Luo, E.~Adeli, Y.~Wang, L.~Lu, A.~L. Yuille, and
  Y.~Zhou, ``{TransUNet: Transformers Make Strong Encoders for Medical Image
  Segmentation},'' \emph{{CoRR}}, vol. abs/2102.04306, 2021. [Online].
  Available: \url{https://arxiv.org/abs/2102.04306}
\BIBentrySTDinterwordspacing

\bibitem{msrf}
\BIBentryALTinterwordspacing
A.~Srivastava, D.~Jha, S.~Chanda, U.~Pal, H.~D. Johansen, D.~Johansen, M.~A.
  Riegler, S.~Ali, and P.~Halvorsen, ``{MSRF-Net: A Multi-Scale Residual Fusion
  Network for Biomedical Image Segmentation},'' \emph{{IEEE J. Biomed. Health
  Informatics}}, vol.~26, no.~5, pp. 2252--2263, 2022. [Online]. Available:
  \url{https://ieeexplore.ieee.org/document/9662196/}
\BIBentrySTDinterwordspacing

\bibitem{vgg19}
\BIBentryALTinterwordspacing
K.~Simonyan and A.~Zisserman, ``{Very Deep Convolutional Networks for
  Large-Scale Image Recognition},'' in \emph{3rd International Conference on
  Learning Representations, {ICLR} 2015, San Diego, CA, USA, May 7-9, 2015,
  Conference Track Proceedings}, Y.~Bengio and Y.~LeCun, Eds., 2015. [Online].
  Available: \url{http://arxiv.org/abs/1409.1556}
\BIBentrySTDinterwordspacing

\bibitem{se}
\BIBentryALTinterwordspacing
J.~Hu, L.~Shen, and G.~Sun, ``{Squeeze-and-Excitation Networks},'' in
  \emph{2018 IEEE/CVF Conference on Computer Vision and Pattern
  Recognition}.\hskip 1em plus 0.5em minus 0.4em\relax {IEEE}, 2018, pp.
  7132--7141. [Online]. Available:
  \url{https://ieeexplore.ieee.org/document/8578843}
\BIBentrySTDinterwordspacing

\bibitem{sog}
G.~D. Finlayson and E.~Trezzi, ``{Shades of Gray and Colour Constancy},'' in
  \emph{The Twelfth Color Imaging Conference: Color Science and Engineering
  Systems, Technologies, Applications, {CIC} 2004, Scottsdale, Arizona, USA,
  November 9-12, 2004}.\hskip 1em plus 0.5em minus 0.4em\relax {IS\&}T - The
  Society for Imaging Science and Technology, 2004.

\bibitem{codella}
\BIBentryALTinterwordspacing
N.~C.~F. Codella, V.~Rotemberg, P.~Tschandl, M.~E. Celebi, S.~W. Dusza, D.~A.
  Gutman, B.~Helba, A.~Kalloo, K.~Liopyris, M.~A. Marchetti, H.~Kittler, and
  A.~Halpern, ``Skin lesion analysis toward melanoma detection 2018: {A}
  challenge hosted by the international skin imaging collaboration {(ISIC)},''
  \emph{CoRR}, vol. abs/1902.03368, 2019. [Online]. Available:
  \url{http://arxiv.org/abs/1902.03368}
\BIBentrySTDinterwordspacing

\bibitem{CVCClinicDB1}
\BIBentryALTinterwordspacing
J.~Bernal, F.~J. S{\'{a}}nchez, G.~Fern{\'{a}}ndez{-}Esparrach, D.~Gil, C.~R.
  de~Miguel, and F.~Vilari{\~{n}}o, ``{WM-DOVA maps for accurate polyp
  highlighting in colonoscopy: Validation vs. saliency maps from physicians},''
  \emph{{Comput. Medical Imaging Graph.}}, vol.~43, pp. 99--111, 2015.
  [Online]. Available:
  \url{https://www.sciencedirect.com/science/article/abs/pii/S0895611115000567}
\BIBentrySTDinterwordspacing

\bibitem{isic2018}
\BIBentryALTinterwordspacing
N.~C.~F. Codella, D.~A. Gutman, M.~E. Celebi, B.~Helba, M.~A. Marchetti, S.~W.
  Dusza, A.~Kalloo, K.~Liopyris, N.~K. Mishra, H.~Kittler, and A.~Halpern,
  ``{Skin Lesion Analysis Toward Melanoma Detection: A Challenge at the 2017
  International Symposium on Biomedical Imaging (ISBI), Hosted by the
  International Skin Imaging Collaboration (ISIC)},'' vol. abs/1710.05006,
  2017. [Online]. Available: \url{http://arxiv.org/abs/1710.05006}
\BIBentrySTDinterwordspacing

\bibitem{DataScienceBowl1}
\BIBentryALTinterwordspacing
J.~C. Caicedo, A.~Goodman, K.~W. Karhohs, B.~A. Cimini, J.~Ackerman,
  M.~Haghighi, C.~Heng, T.~Becker, M.~Doan, C.~McQuin \emph{et~al.}, ``{Nucleus
  segmentation across imaging experiments: the 2018 Data Science Bowl},''
  \emph{{Nature methods}}, vol.~16, no.~12, pp. 1247--1253, 2019. [Online].
  Available: \url{https://www.nature.com/articles/s41592-019-0612-7}
\BIBentrySTDinterwordspacing

\bibitem{sudre2017generalised}
\BIBentryALTinterwordspacing
C.~H. Sudre, W.~Li, T.~Vercauteren, S.~Ourselin, and M.~J. Cardoso,
  ``Generalised dice overlap as a deep learning loss function for highly
  unbalanced segmentations,'' in \emph{Deep Learning in Medical Image Analysis
  and Multimodal Learning for Clinical Decision Support - Third International
  Workshop, {DLMIA} 2017, and 7th International Workshop, {ML-CDS} 2017, Held
  in Conjunction with {MICCAI} 2017, Qu{\'{e}}bec City, QC, Canada, September
  14, 2017, Proceedings}, ser. Lecture Notes in Computer Science, M.~J.
  Cardoso, T.~Arbel, G.~Carneiro, T.~F. Syeda{-}Mahmood, J.~M. R.~S. Tavares,
  M.~Moradi, A.~P. Bradley, H.~Greenspan, J.~P. Papa, A.~Madabhushi, J.~C.
  Nascimento, J.~S. Cardoso, V.~Belagiannis, and Z.~Lu, Eds., vol. 10553.\hskip
  1em plus 0.5em minus 0.4em\relax Springer, 2017, pp. 240--248. [Online].
  Available: \url{https://doi.org/10.1007/978-3-319-67558-9\_28}
\BIBentrySTDinterwordspacing

\bibitem{dozat}
T.~Dozat, ``{Incorporating Nesterov Momentum into Adam},'' in
  \emph{International Conference on Learning Representations, Caribe Hilton,
  San Juan, Puerto Rico, May 2 - 4, 2016}, 2016.

\bibitem{album}
\BIBentryALTinterwordspacing
A.~Buslaev, V.~I. Iglovikov, E.~Khvedchenya, A.~Parinov, M.~Druzhinin, and
  A.~A. Kalinin, ``{Albumentations: Fast and Flexible Image Augmentations},''
  \emph{{Inf.}}, vol.~11, no.~2, p. 125, 2020. [Online]. Available:
  \url{https://www.mdpi.com/2078-2489/11/2/125}
\BIBentrySTDinterwordspacing

\bibitem{zou2004statistical}
\BIBentryALTinterwordspacing
K.~H. Zou, S.~K. Warfield, A.~Bharatha, C.~M. Tempany, M.~R. Kaus, S.~J. Haker,
  W.~M. Wells~III, F.~A. Jolesz, and R.~Kikinis, ``{Statistical validation of
  image segmentation quality based on a spatial overlap index1: scientific
  reports},'' \emph{{Academic Radiology}}, vol.~11, no.~2, pp. 178--189, 2004.
  [Online]. Available:
  \url{https://www.sciencedirect.com/science/article/pii/S1076633203006718}
\BIBentrySTDinterwordspacing

\bibitem{deeplab}
\BIBentryALTinterwordspacing
L.~Chen, Y.~Zhu, G.~Papandreou, F.~Schroff, and H.~Adam, ``{Encoder-Decoder
  with Atrous Separable Convolution for Semantic Image Segmentation},'' in
  \emph{Computer Vision - {ECCV} 2018 - 15th European Conference, Munich,
  Germany, September 8-14, 2018, Proceedings, Part {VII}}, ser. Lecture Notes
  in Computer Science, V.~Ferrari, M.~Hebert, C.~Sminchisescu, and Y.~Weiss,
  Eds., vol. 11211.\hskip 1em plus 0.5em minus 0.4em\relax {Springer}, 2018,
  pp. 833--851. [Online]. Available:
  \url{https://link.springer.com/chapter/10.1007/978-3-030-01234-2_49}
\BIBentrySTDinterwordspacing

\bibitem{res-unet++}
\BIBentryALTinterwordspacing
D.~Jha, P.~H. Smedsrud, M.~A. Riegler, D.~Johansen, T.~de~Lange, P.~Halvorsen,
  and H.~D. Johansen, ``{ResUNet++: An Advanced Architecture for Medical Image
  Segmentation},'' in \emph{{IEEE} International Symposium on Multimedia, {ISM}
  2019, San Diego, CA, USA, December 9-11, 2019}.\hskip 1em plus 0.5em minus
  0.4em\relax {IEEE}, 2019, pp. 225--230. [Online]. Available:
  \url{https://ieeexplore.ieee.org/document/8959021/}
\BIBentrySTDinterwordspacing

\bibitem{uacanet}
\BIBentryALTinterwordspacing
T.~Kim, H.~Lee, and D.~Kim, ``{UACANet: Uncertainty Augmented Context Attention
  for Polyp Segmentation},'' in \emph{{MM} '21: {ACM} Multimedia Conference,
  Virtual Event, China, October 20 - 24, 2021}, H.~T. Shen, Y.~Zhuang, J.~R.
  Smith, Y.~Yang, P.~C{\'{e}}sar, F.~Metze, and B.~Prabhakaran, Eds.\hskip 1em
  plus 0.5em minus 0.4em\relax {ACM}, 2021, pp. 2167--2175. [Online].
  Available: \url{https://dl.acm.org/doi/10.1145/3474085.3475375}
\BIBentrySTDinterwordspacing

\bibitem{swin}
\BIBentryALTinterwordspacing
Z.~Liu, Y.~Lin, Y.~Cao, H.~Hu, Y.~Wei, Z.~Zhang, S.~Lin, and B.~Guo, ``{Swin
  Transformer: Hierarchical Vision Transformer using Shifted Windows},'' in
  \emph{2021 {IEEE/CVF} International Conference on Computer Vision, {ICCV}
  2021, Montreal, QC, Canada, October 10-17, 2021}.\hskip 1em plus 0.5em minus
  0.4em\relax {IEEE}, 2021, pp. 9992--10\,002. [Online]. Available:
  \url{https://ieeexplore.ieee.org/document/9710580/}
\BIBentrySTDinterwordspacing

\end{thebibliography}

\end{document}